\documentclass{mn2e}

\usepackage{amsfonts}
\usepackage{amsmath}
\usepackage{graphicx}

\title[Photo-$z$ redshift and luminosity distributions]
{On estimating redshift and luminosity distributions in 
 photometric redshift surveys}

\author[R. K. Sheth]
{Ravi K. Sheth\thanks{E-mail:  shethrk@physics.upenn.edu}\\
 Department of Physics \& Astronomy, University of Pennsylvania, 
 209 S. 33rd Street, Philadelphia, PA 19104, USA}

\newcommand{\bm}[1]{{\mbox{\boldmath $#1$}}}

\begin{document}
\pagerange{\pageref{firstpage}--\pageref{lastpage}}

\maketitle

\label{firstpage}

\begin{abstract}
The luminosity functions of galaxies and quasars provide invaluable 
information about galaxy and quasar formation.  Estimating the 
luminosity function from magnitude limited samples is relatively 
straightforward, provided that the distances to the objects in the 
sample are known accurately; techniques for doing this have been 
available for about thirty years.  
However, distances are usually known accurately for only a small subset 
of the sample.  This is true of the objects in the Sloan Digital Sky 
Survey, and will be increasingly true of the next generation of deep 
multi-color photometric surveys.  
Estimating the luminosity function when distances are only known 
approximately (e.g., photometric redshifts are available, but 
spectroscopic redshifts are not) is more difficult.  I describe two 
algorithms which can handle this complication:  one is a generalization 
of the $V_{\rm max}$ algorithm, and the other is a maximum likelihood 
approach.  Because these methods account for uncertainties in the 
distance estimate, they impact a broader range of studies.  
For example, they are useful for studying the abundances of galaxies 
which are sufficiently nearby that the contribution of peculiar 
velocity to the spectroscopic redshift is not negligible, so only a 
noisy estimate of the true distance is available.  In this respect, 
peculiar velocities and photometric redshift errors have similar effects.  
The methods developed here are also useful for estimating the 
stellar luminosity function in samples where accurate parallax 
distances are not available.  
\end{abstract}

\begin{keywords}
methods: analytical - galaxies: formation - galaxies: haloes -
dark matter - large scale structure of the universe 
\end{keywords}

\section{Introduction}
Estimates of the distribution of distances to galaxies, and of the 
galaxy luminosity function and its evolution, provide useful constraints 
on models of galaxy formation.  
Current (e.g. the SDSS, York et al. 2000, 
              Combo-17, Wolf et al. 2003, 
              MUSYC, Marchesini et al. 2007) 
and planned surveys (e.g., DES, LSST) go considerably deeper in 
multicolor photometry than in spectroscopy, or are entirely photometric.  
For such surveys, reasonably accurate photometric redshift estimates 
(e.g. Hyper-$z$, Bolzonella et al. 2000, and ANN$z$, 
Collister \& Lahav 2003) can or will be made.  
In the case of Luminous Red Galaxies (e.g. Eisenstein et al. 2001), 
the photometric redshifts may actually be quite accurate (e.g. 
Padmanabhan et al. 2004; Weinstein et al. 2004; Collister et al. 2006).  
The number of objects with photometric redshifts typically exceeds the 
number for which spectroscopic redshifts are available by more than an 
order of magnitude.  
This is also true of new quasar detection algorithms.  Whereas the 
SDSS will obtain spectra of about one hundred thousand quasars, the 
Non-parametric Bayesian Classification algorithm of Richards et al. (2004) 
has identified one million quasars using SDSS photometry.  Large 
photometric samples of galaxies and quasars offer the potential of 
studying cosmological evolution at a fraction of the cost of a full 
spectroscopic survey.  

Bigger is not better only for studying the evolution of the galaxy and 
quasar populations.  In the case of galaxies, the larger number of 
LRGs with photometric redshifts, allowed new science:  
the detection of the integrated Sachs-Wolfe effect (Fosalba et al. 2003; 
Scranton et al. 2003; Padmanabhan et al. 2005; Cabre et al. 2006) 
required the larger photometric LRG catalog.  
In the case of quasars also, larger sample sizes allow one to address 
new science questions.  For example, the SDSS spectroscopic sample is 
barely large enough to measure the gravitational lensing magnification 
bias signal with high statistical significance:  the larger photometric 
sample made the measurement possible (Scranton et al. 2005).  

With photometric redshift surveys becoming the norm, it is timely to 
devise methods for estimating the distribution of comoving distances 
and the evolution of the luminosity function in such samples.  
Broadly speaking, techniques for estimating the luminosity function 
from a magnitude limited catalog fall into two classes:  one is based 
on the nonparametric $V_{max}$ method outlined by Schmidt (1968); 
the other is a maximum likelihood analysis which can provide parametric 
or nonparametric estimates of the luminosity function (Sandage, 
Tammann \& Yahil 1979; Efstathiou, Ellis \& Peterson 1988; 
Springel \& White 1998).  
Both methods assume that the distances are known precisely and 
accurately.  
The main goal of the present work is to generalize both types of 
methods to handle photometric redshifts.  
For reasons described below, the analysis which follows is 
best-suited to studying objects where evolution and $k$-correction 
uncertainties are small.  In practice, this means they are best 
suited to catalogs which contain objects of one spectral type.  
Removing this constraint is the subject of ongoing work.  

Section~\ref{vmax} discusses a deconvolution algorithm for 
estimating $dN/dz$ and the luminosity function from photometric 
redshift samples.  The estimator of the luminosity function is a 
generalization of  the $V_{\rm max}$ method (Schmidt 1968), and 
the method uses the deconvolution algorithm described by Lucy (1974).  
Section~\ref{ml} discusses a maximum likelihood approach.  
Some applications are discussed in Section~\ref{apps} and 
a final section summarizes.

\section{The $V_{\rm max}$ method}\label{vmax} 
I first outline 
why the problems of estimating $dN/dz$ and $\phi(L)$ are both 
best thought of as deconvolution problems.  I then show that 
Lucy's deconvolution algorithm provides an efficient way of 
performing the deconvolution.  

\subsection{The redshift distribution:  $dN/dz$}
Let $dN/dz$ denote the number of objects which lie at redshift $z$ 
(since peculiar velocities are unlikely to be larger than a few 
thousand km/s, they do not make a significant change to the redshift 
if $z>0.01$).  Let $p(z_e|z)$ denote the probability of estimating the 
redshift as $z_e$ when the true value is $z$.  Then the distribution 
of estimated redshifts is 
\begin{equation}
 {dN_e(z_e)\over dz_e} = \int dz\, {dN(z)\over dz}\,p(z_e|z).
\end{equation}
To get an idea of what this implies, suppose that $p(z_e|z)$ is sharply 
peaked around the true value $z$.  Then define $\Delta z\equiv z_e - z$ 
and expand $dN/dz$ in a Taylor series around its 
value at $z_e$.  This yields an expansion in $\Delta z$.  If the estimated 
redshift is unbiased in the mean, then $\langle\Delta z\rangle=0$ and the 
leading order contribution is of the form 
\begin{equation}
 {dN_e(z_e)\over dz_e} \approx 
 {dN(z_e)\over dz_e} + {\langle\Delta z^2\rangle\over 2}\,
                         {\partial^2 [dN(z)/dz]\over\partial z^2}\Biggl|_{z_e}
\end{equation}
Typically, $dN/dz$ is well approximated by a constant times 
$z^2\,\exp[-(z/z_m)^\alpha]$, with $\alpha\approx 3/2$ and $z_m$ set by 
the luminosity function and the limiting magnitude of the catalog 
(i.e., $dN/dz\propto z^2$ at $z\ll z_m$, and it drops rapidly for 
$z\gg z_m$).  In this case, 
\begin{equation}
 {dN_e(z_e)\over dz_e} \approx 
 {dN(z_e)\over dz_e}\left[1 + {\langle\Delta z^2\rangle\over z_e^2}\,
  C(z_e)\right],
\end{equation}
where 
\begin{equation}
 C(z_e) = 1 - {3\alpha\over 2}\left({z_e\over z_*}\right)^\alpha
      - {\alpha^2\over 2} \left({z_e\over z_*}\right)^\alpha
      + {\alpha^2\over 2} \left({z_e\over z_*}\right)^{2\alpha}.
\end{equation}
The term in square brackets shows how the estimated distribution 
$dN_e$ differs from the true one $dN$.  In particular, it shows that 
an accurate estimate of $dN$ can be obtained by summing over 
all objects that have estimated redshift $z_e$, weighting each by 
the inverse of the term in square brackets in the expression above.

The general problem is to infer the shape of the intrinsic distribution 
$dN/dz$ given the measured distribution $dN_e/dz_e$, even if $p(z_e|z)$ 
is not sharply peaked.  If $p(z_e|z)$ is known, and $dN_e/dz_e$ is 
measured, then this is an integral equation of the first kind, which 
can be solved to obtain the intrinsic $dN/dz$.  This is possible even 
if $p(z_e|z)$ is fairly broad.  
Padmanabhan et al. (2004) describe a method to do this, but, for 
reasons made explicit in Lucy (1974), their method is not ideal.  
Before we describe our method, the following section shows that 
estimating the intrinsic luminosity function from photometric 
redshift data is a similar deconvolution problem.  

\subsection{The luminosity distribution:  $\phi(L)$}
Let $\phi(M|z)$ denote the number density of galaxies with absolute 
magnitudes $M\propto -2.5\log_{10}L$, where $L=\ell\,4\pi D_L^2(z)$ is the 
luminosity, $\ell$ is the apparent brightness, and $D_L(z)$ is the 
luminosity distance at $z$.  Assume for the moment that there is no 
evolution (extending the analysis to include evolution is the subject 
of work in progress).  Then $\phi(M|z)$ is independent of $z$.  

Simply adding up the total number of galaxies in a magnitude limited 
catalog which have luminosity $L$ and dividing by the total volume of 
the survey is not a good estimator of $\phi(L)$ itself.  This is because 
the more luminous objects will be visible to larger distances.  
Let $V_{\rm max}(M)$ denote the largest comoving volume out to which an 
object of absolute magnitude $M$ can be seen.  If the catalog is limited 
at both ends, then there is a minimum volume below which the object 
would have been too bright to be included in the catalog:  call this 
$V_{min}(M)$.  The number of galaxies with absolute magnitude $M$ in a 
catalog magnitude limited at both ends is  
\begin{equation}
 N(M) = \phi(M)\, \Bigl[V_{\rm max}(M)-V_{\rm min}(M)\Bigr].
 \label{NMVmax}
\end{equation}
Therefore, if we sum over all the galaxies in a magnitude limited catalog, 
and we weight each object by the inverse of 
$V_{\rm max}(M)-V_{\rm min}(M)$, then we will actually have estimated 
the luminosity function.  This is the basis of the $1/V_{\rm max}$ 
method (Schmidt 1968).  

If the estimate $z_e$ of the true redshift $z$ comes with a large 
uncertainty, this translates into an uncertainty in the luminosity 
(this assumes that the error in redshift determination does not affect 
the observed apparent magnitude).  
The total number of objects with estimated absolute magnitudes $M_e$ is 
\begin{eqnarray}
 N_e(M_e) 
  &=& \int_{m_{\rm min}}^{m_{\rm max}} {\rm d}m\,
       \int {\rm d}D_{\rm L}\, n(m,D_{\rm L})\, 
       p(m-M_e|D_{\rm L},m)  \nonumber\\
  &=& \int {\rm d}D_{\rm L} {{\rm d}V_{\rm com}\over {\rm d}D_{\rm L}}
         \int_{M_{\rm min}(D_{\rm L})}^{M_{\rm max}(D_{\rm L})} 
           {\rm d}M\,\phi(M)\,\nonumber\\
  & & \qquad\qquad\qquad\qquad \times\quad p(M-M_e|D_{\rm L},M) \nonumber\\
  &=& \int {\rm d}M\,\phi(M) 
      \int_{D_{\rm L}(M_{\rm min})}^{D_{\rm L}(M_{\rm max})} {\rm d}D_{\rm L}
            {{\rm d}V_{\rm com}(D_{\rm L})\over {\rm d}D_{\rm L}} \nonumber\\
  & & \qquad\qquad \times\quad p(M-M_e|D_{\rm L},M),
 \end{eqnarray}
where we have used the fact that 
 $5\log_{10}D_e = m - M_e = 5\log_{10}D_{\rm L}+M-M_e$, so 
$p(D_e|D_{\rm L},m)\,{\rm d}D_e = p(M-M_e|D_{\rm L},m)\,{\rm d}M_e$.  
Note that if there is no error in the distance, then $p(M-M_e)$ is 
a delta function centered on $M$, and this expression reduces to 
equation~(\ref{NMVmax}).  

\begin{figure*}
\centering
 \includegraphics[width=0.92\hsize]{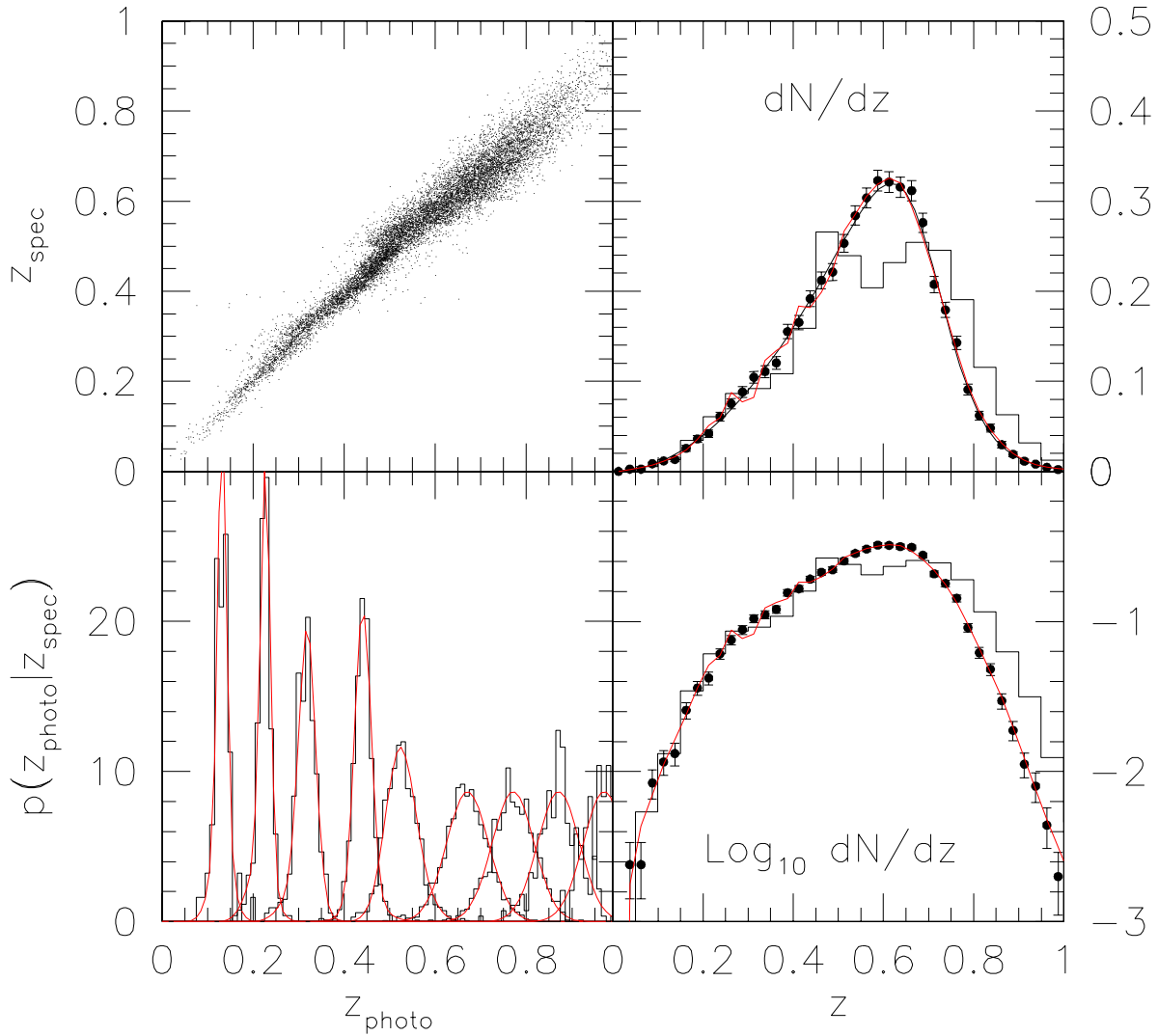}
 \vspace{-1cm}
 \caption{Example of the difference between the intrinsic redshift 
         distribution (filled circles with error bars in panels on 
         right) and the photometric redshift distribution (histograms 
         in panels on right).  Panels on the left compare the 
         intrinsic and estimated redshifts.  Jagged lines in panels 
         on right show how successive iterations converge rapidly to 
         the intrinsic distribution:  the histogram was used as the 
         starting guess.  }
\label{NZlucy}
\end{figure*}

If ${\cal V}(V_{\rm max},V_{\rm min},M)$ denotes  result of 
performing the integral over $D_{\rm L}$ in the final expression 
above, then 
\begin{equation}
 N_e(M_e) 
  = \int {\rm d}M\,\phi(M) \,{\cal V}(V_{\rm max},V_{\rm min},M).
 \label{NeMe}
\end{equation}
Since $V_{\rm max}$ and $V_{\rm min}$ are known functions of $M$, 
${\cal V}$ itself is really just a complicated function of $M$.  
To get some feel for its form, suppose that the error in determining 
the redshift does not depend on apparent magnitude, and, in addition, 
the error distribution is a function of the ratio $D_e/D_{\rm L}$ 
only.  Then $p(M-M_e|D_{\rm L},M)$ does not depend explicitly on 
$D_{\rm L}$ itself, so it can be taken out of the integral over 
$D_{\rm L}$.  In this case, 
\begin{equation}
 {\cal V} = \Bigl[V_{\rm max}(M)-V_{\rm min}(M)\Bigr]\,p(M_e|M), 
\end{equation}
Now, $\phi(M)$ times the term in square brackets is the intrinsic 
$N(M)$ distribution (equation~\ref{NMVmax}), so equation~(\ref{NeMe}) 
becomes 
\begin{equation}
 N_e(M_e) = \int {\rm d}M\,N(M)\,p(M_e|M).
\end{equation}
In this case, the observed distribution of $M_e$ is the convolution, 
not just of the luminosity function $\phi$ with the error distribution 
$p$, but of the product of $\phi$ and $(V_{\rm max}-V_{\rm min})$ 
with $p$.  The inclusion of this second term accounts for the fact that 
more objects are likely to scatter down from large $z$ to small than 
the other way around, simply because there is a greater volume at 
larger $z$.  The form of the expression above shows clearly that one 
generically expects distance errors to scatter objects from the 
peak of the $N(M)$ distribution to the tails.  Unless it is corrected 
for, this will lead one to overestimate the number density of low 
and high luminosity objects relative to the mean.  

When the distances are known accurately, one can simply use 
$\phi = N/V$ as a non-parametric estimate of the luminosity function.  
However, the expression above shows that the relation between 
$N_e(M_e)$ and $\phi$ is more complicated than when the distance is 
not known precisely:  determining $\phi$ requires solution of an 
integral equation.  In this respect, the problem is similar to that 
of determining $dN/dz$ when $dN_e/dz_e$ and $p(z_e|z)$ are known.  
Once again, in the case of small errors, one can expand the integrand 
in a power series and then perform the integral to determine the 
correction factor $C(M_e)$ that is required if one wishes to weight 
galaxies by $1/[(1+C)V]$ and so estimate $\phi$ from the number of 
observed $M_e$.  But the general case is more complicated.  

Before moving on to the solution, note that the assumption that 
$p(M-M_e|D_{\rm L},M)$ does not depend explicitly on $D_{\rm L}$ 
itself, is not crucial.  I have mainly made the assumption here 
so that the form of the argument is clear.  If it does depend on 
$D_{\rm L}$, then the weighting factor in the integrand is a more 
complicated function of $M$ than simply $N(M)\,p(M_e|M)$.  

\subsection{Non-parametric deconvolution and the $V_{\rm max}$ method}
Since $N_e(M_e)$, $p(M_e|M)$ and ${\rm d}V_{\rm com}/{\rm d}z$ are 
all known, the relation to be solved for $\phi(M)$ is an integral 
equation of the first kind.  Standard arguments show that it can be 
written as a matrix equation which can then be solved for $\phi(M)$.  
The problem with this approach is how one accounts for the fact that 
the measured $N_e(M_e)$ distribution may be noisy.  
In particular, since $N_e(M_e)$ is likely to be smoother than $N(M)$, 
if $N_e$ contains sharp features, then the recovered $N$ will contain 
sharper features.  If sharp features are expected to be unrealistic, 
and the measurement is noisy (this will always be true in the tails), 
then an exact inversion of the integral equation is clearly undesirable.  
An iterative algorithm which avoids this problem was proposed by 
Lucy (1974); it converges rapidly and is simple to code ($\sim 20$ 
lines of code), so it is the method of choice.  

\begin{figure}
\centering
 \vspace{-4.7cm}
 \includegraphics[width=\hsize]{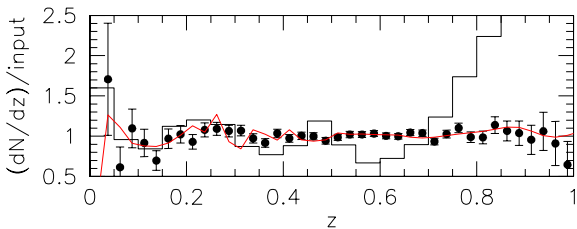}
 \vspace{-1cm}
 \caption{Comparison of true intrinsic redshift distribution and 
          that recovered by the algorithm described in the text.  
          Symbols with error bars show one realization of the 
          intrinsic distribution (the difference from unity is 
          `shot-noise' due to the finite size of the sample).  
          Histogram shows the associated photo-$z$ distribution, 
          and jagged curve shows the recovered distribution after 
          four iterations:  the histogram was used as the 
          starting guess.}  
\label{NZratio}
\end{figure}

Figure~\ref{NZlucy} shows how well this method works on mock data.  
Mock galaxies were distributed in redshift as indicated by the 
filled circles in the right-hand panels of Figure~\ref{NZlucy}.  
Estimated redshifts were assigned as shown in the bottom left 
panel (the particular choice of $p(z_e|z)$ will be discussed 
shortly).  Top left panel compares the estimated and true redshifts.  
The histograms in the panels on the right show the distribution of 
estimated redshifts.  Note how different they are from the true 
distribution:  although $dN/dz$ has a single well-defined peak, 
$dN_e/dz_e$ is almost bimodal.  Our choice of $p(z_e|z)$ was chosen 
to produce just this effect:  it mimics the effect on some photometric 
redshift estimators as, e.g., the 4000\AA\ break passes from one filter 
to another.  The problem is to use the estimated histogram and the 
known shape of $p(z_e|z)$ to infer that the true intrinsic distribution 
traces the locus defined by the filled circles.  
The histogram was used as the starting guess for the deconvolution 
algorithm, after which the algorithm converged rapidly to the filled 
circles (four iterations are shown; they overlap one another closely).
Figure~\ref{NZratio} provides a more detailed comparison of how 
well the recovered distribution resembles the true one, and how 
different the photo-$z$ distribution, which was used as the starting 
guess, is from the true distribution.  

\begin{figure*}
 \centering
 \includegraphics[width=\hsize]{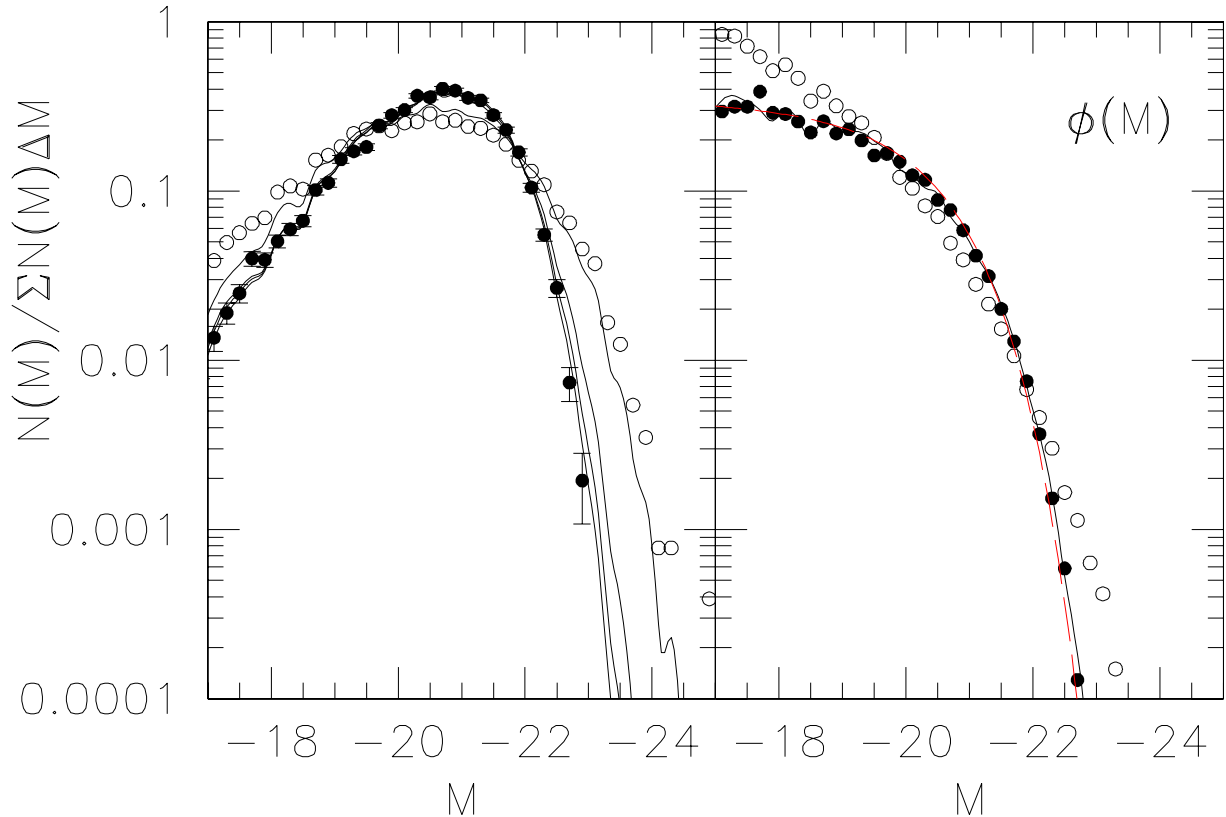}
 \vspace{-5.5cm}
 \caption{Reconstruction of the intrinsic $N(M)$ distribution (filled 
          circles) from the distribution of estimated redshifts when 
          distance uncertainties are large (open circles).  Error bars 
          on the filled circles assume Poisson statistics.  Different 
          curves show how successive iterations of the deconvolution 
          algorithm approximate the intrinsic distribution increasingly 
          well:  the open circles were used as the starting guess, and 
          curves show results after iterations 1, 8, 15, and 22.  
          Panel on the left shows the observed distribution, and 
          panel on the right shows the associated estimate of the 
          luminosity function.  Dashed curve shows the input 
          luminosity function.  The generic effect of photo-$z$ 
          errors, which the deconvolution algorithm rectifies 
          (compare solid line with filled circles), 
          is to enhance the large luminosity tail, and steepen 
          the faint end slope (compare open with filled circles).}
 \label{NMvpec}
\end{figure*}

Figure~\ref{NMvpec} shows results for the luminosity function.  
The panel on the left shows the intrinsic $N(M)$ (solid circles) 
and estimated $N_e(M)$ (open circles) distributions in a mock 
catalog generated assuming the same flat cosmological model as 
before, but with the intrinsic distribution of luminosities and 
the apparent magnitude limits chosen to be those of the galaxies in 
the SDSS survey (Blanton et al. 2003).  
The estimated redshifts were assumed to follow 
$p(D_e|D)dD_e = (dx/x) (\gamma x)^\gamma\exp(-\gamma x)/\Gamma(\gamma)$, 
where $x=D_e/D$ and $\gamma=5$.  This distribution has 
$\langle x\rangle = 1$, and $\sigma^2_x = 1/\gamma$.  With $\gamma=5$, 
this error distribution is substantially worse than typical photometric 
redshift errors.  Notice how $N(M_e)$ is broader than the true 
distribution:  it has noticably more objects in the tails, and hence 
fewer near the peak.  This is the generic effect we mentioned earlier. 

The open circles in the panel on the right show the result of 
converting from $N_e(M)$ to $\phi(M)$ using Schmidt's method with 
no correction for the photometric redshift error distribution.  This 
estimate has more luminous galaxies, and a steeper faint-end slope, 
than the true distribution shown by the solid circles.  For photo-$z$ 
error distributions which are approximately symmetric, this sort of 
discrepancy is generic.  

The solid lines in the panel on the left show successive iterations 
of the deconvolution algorithm, starting from the open circles.  
Convergence to the correct distribution is clearly seen.  The solid 
line in the panel on the left shows the result of applying 
Schmidt's method to the estimate of $N(M)$ returned by the 
final iteration shown.  It is an excellent approximation to the 
intrinsic distribution.

To illustrate that the generic effect of distance errors is to scatter 
objects from the peak of $N(M_e)$ into the tails, thus increasing the 
expected number of high luminosity objects and increasing the slope at 
small luminosities, Figure~\ref{NMgaussian} shows a similar 
calculation, but now when the true intrinsic luminosity function is a 
Gaussian in absolute magnitude.  The precise parameter values were 
chosen to match those of early-type galaxies in the SDSS, and, once 
again, I have assumed photo-$z$ error distributions 
(Gaussian with 1~mag rms) that are significantly broader than most 
photo-$z$ algorithms return.  Notice again how the algorithm rapidly 
converges from the observed counts (open circles) to the true ones 
(filled circles).  

\section{The maximum-likelihood method}\label{ml}
In magnitude limited samples, an unbiased estimate of the luminosity 
function is obtained by maximizing the (log of the) likelihood function 
\begin{eqnarray}
 {\cal L}({\bm a}) &=& \prod_i p_i, \quad {\rm where}\quad \nonumber\\
 p_i &=& {\phi(L_i|z_i,{\bm a})\over 
      \int_{L_{\rm min}(z_i)}^{L_{\rm max}(z_i)}{\rm d}L\,\phi(L|z_i,{\bm a})} 
     = {\phi(L_i|z_i,{\bm a})\over S(z_i,{\bm a})} 
 \label{mlspecz}
\end{eqnarray}
(Sandage, Tammann \& Yahil 1979; Efstathiou, Ellis \& Peterson 1988).  
Here $z_i$ denotes the redshift of galaxy $i$, 
$\phi(L|z,{\bm a})$ is the luminosity function at $z$, with shape 
specified by the parameters ${\bm a}$, 
and $L_{\rm min}(z)$ is the minimum luminosity which a galaxy at 
$z$ must have to be observed in the flux limited catalog.  
That is to say, the parameters $a$ which specify the luminosity 
function are those for which
\begin{equation}
  {\partial\, {\rm ln}{\cal L}\over \partial a} = 
  \sum_i {\partial\, {\rm ln}\,p_i\over \partial a} = 0.
\end{equation}
Note that our notation allows the model luminosity function to have a 
parametric form, in which case ${\bm a}$ denotes the free parameters 
of the model, or to be non-parametric, in which case the luminosity 
function is represented as a sum over bins in luminosity, and ${\bm a}$ 
denotes the parameters necessary to specify the bin shapes---the most 
popular shapes being tophats, or Gaussians, or concave polynomials with 
compact support.  

If the redshift $z$ is not known precisely, and if the inaccuracy in 
redshift does not affect the observed apparent magnitude, then the 
method should be modified as follows.  
Let $L_i$ and $z_i$ denote the true luminosity and redshift of 
galaxy $i$, which together determine $\ell_i$, the observed apparent 
brightness of the object.  If $\zeta_i$ denotes the estimated redshift, 
then this, with $\ell_i$, determines the estimated luminosity which we 
will denote $\lambda_i$.  

The number of objects in a flux-limited catalog with estimated values 
$\zeta$ and $\lambda$ depends on the true intrinsic distribution of 
$L$ and $z$, and on the distribution of redshift errors.  
Since errors in redshift do not alter the observed apparent brightness, 
the number distribution of estimated redshifts is expected to be 
\begin{eqnarray}
 N(\zeta,{\bm a}) &=& \int {\rm d}z {{\rm d}V_{\rm com}\over {\rm d}z}
  \int_{\ell_{\rm min}}^{\ell_{\rm max}} {\rm d}\ell\, 4\pi D_{\rm L}^2(z)\,
        \nonumber\\
 & & \qquad\qquad\times\quad  
      \phi\Bigl(4\pi D_{\rm L}^2(z)\ell\Big|{\bm a}\Bigr)\,p(\zeta|z,\ell),
\end{eqnarray}
if the intrinsic distribution is parametrized by ${\bm a}$.  
Here $p(\zeta|z,\ell)$ represents the distribution of estimated 
redshifts $\zeta$ given true $z$ and $\ell$.  
Similarly, the joint distribution of estimated $\lambda$ and $\zeta$ is 
\begin{eqnarray}
 \lambda\,N(\lambda,\zeta,{\bm a}) &=& 
  \int {\rm d}z {{\rm d}V_{\rm com}\over {\rm d}z}\,
    4\pi D_{\rm L}^2(z)\ell\,\nonumber\\
  & & \qquad\times\quad
   \phi\Bigl(4\pi D_{\rm L}^2(z)\ell\Big|{\bm a}\Bigr)\, p(\zeta|z,\ell).
\end{eqnarray}
Notice that if the redshift-error distribution is independent of $\ell$, 
then 
\begin{eqnarray}
 N(\zeta,{\bm a}) &=& \int {\rm d}z\,{\rm d}V_{\rm com}/{\rm d}z\,
                     S(z,{\bm a})\,p(\zeta|z) \nonumber\\
                  &\equiv& \int {\rm d}z\,N(z,{\bm a})\,p(\zeta|z):  
\end{eqnarray}
this is just the convolution of the intrinsic redshift distribution 
(in a flux-limited catalog) with the redshift-error distribution.  


By analogy to when the distances are known accurately, 
the likelihood to be maximized is ${\cal L} = \prod_i\, p_i$, 
where $p_i$ is the fraction of the number of objects expected to have 
estimated redshifts $\zeta_i$ which also have estimated luminosity 
$\lambda_i$:  
\begin{equation}
 {\cal L}({\bm a}) = \prod_i\, p_i, \qquad {\rm where} \qquad 
               p_i = {N(\lambda_i,\zeta_i,{\bm a})\over N(\zeta_i,{\bm a})}.  
 \label{mlphotoz}
\end{equation}
This expression for $p_i$ differs from that in the literature 
(Chen et al. 2003 is missing the factors of ${\rm d}V_{\rm com}$ 
in the integrals which define the numerator and denominator).  

\begin{figure*}
 \centering
 \includegraphics[width=0.6\hsize]{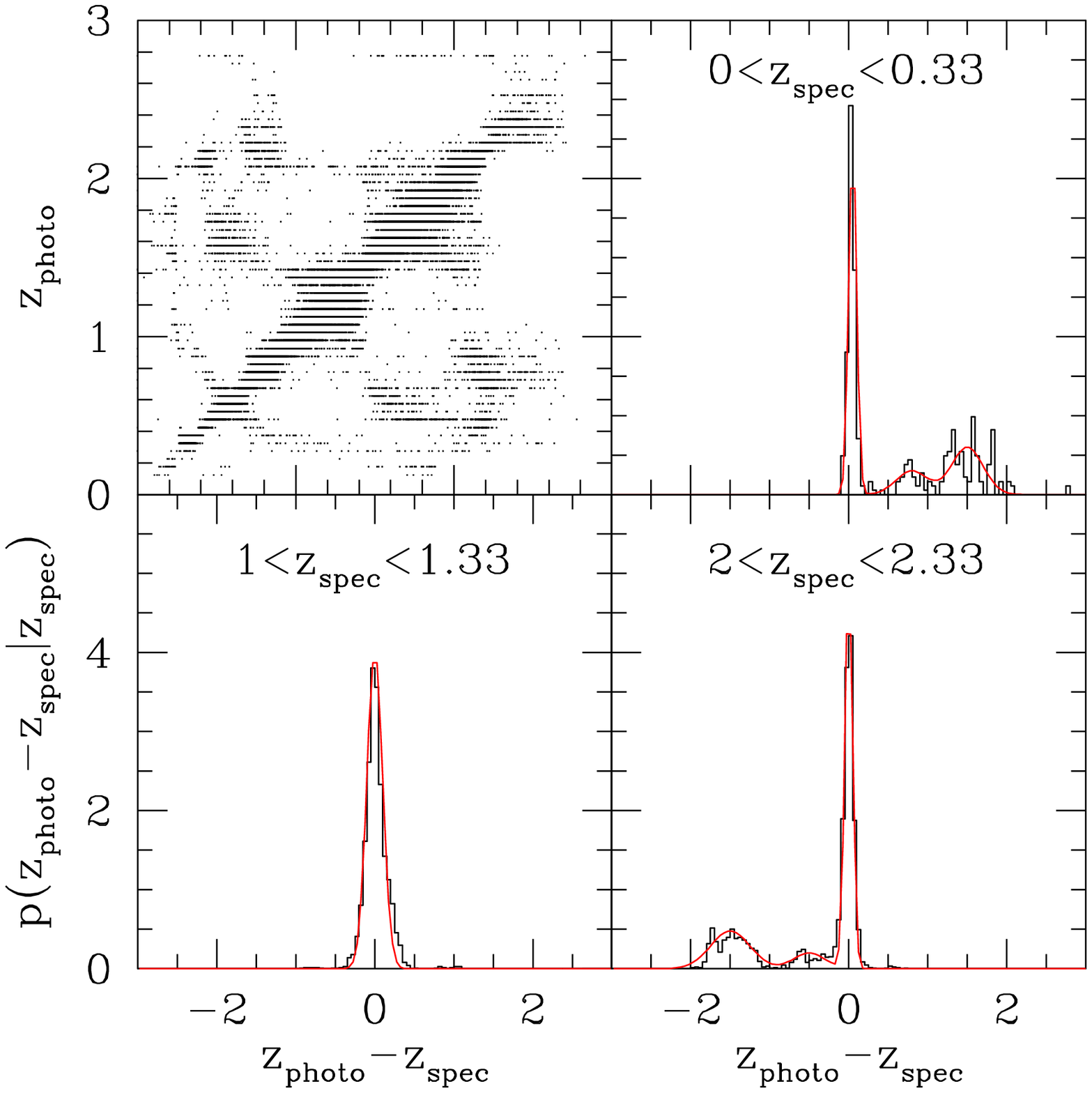}
 \includegraphics[width=0.34\hsize]{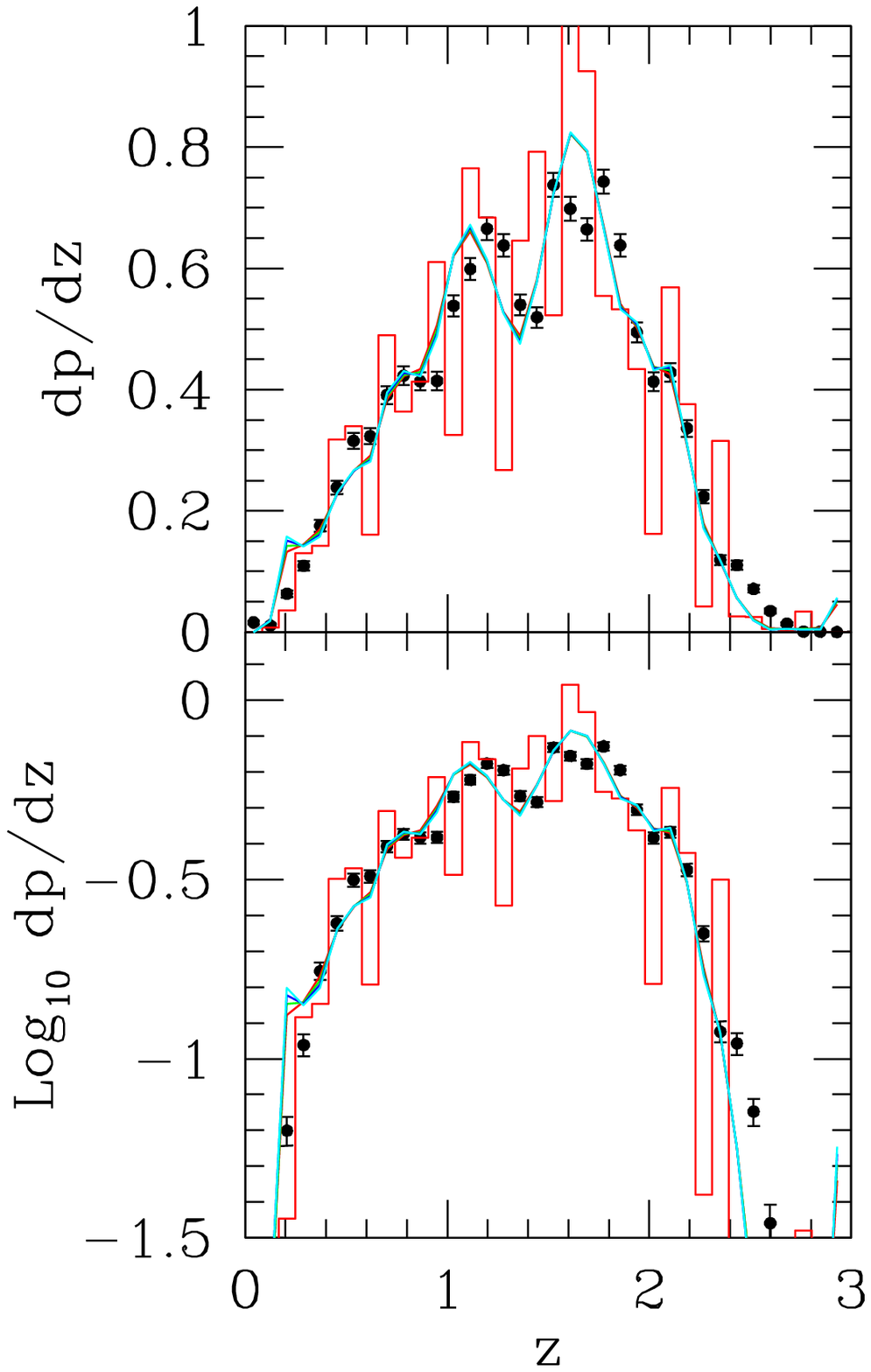}
 \caption{Left:  Distribution of spectroscopic and photometric 
          redshifts in  the SDSS DR1 NBC QSO catalog.  
          Right:  Reconstruction of the intrinsic $dN/dz$ distribution 
          (filled circles) from that of the photometric redshifts 
          (histogram) using the deconvolution algorithm described 
          previously (curve).  The reconstruction is quite accurate, 
          despite the complicated nature of the distance errors.}
 \label{nbc}
\end{figure*}

To check that this expression is indeed the correct one, note that 
maximizing the (log of the) likelihood requires evaluation of 
 $\sum_i \partial\, {\rm ln}p_i/\partial a$.  
This reduces to taking the difference of two terms, the first of which is 
\begin{displaymath}
 \sum_i {\partial\,{\rm ln} N_e(\lambda_i,\zeta_i,{\bm a})\over\partial a} 
  \to   \int {\rm d}\zeta\,\int {\rm d}\lambda 
             {N_t(\lambda,\zeta)\over N_e(\lambda,\zeta,{\bm a})} 
             {\partial N_e(\lambda,\zeta,{\bm a})\over\partial a} ,
\end{displaymath}
where we have written the sum over objects as an integral over their 
estimated redshifts and luminosities.  Similarly, the second term is 
\begin{displaymath}
 \sum_i {\partial\,{\rm ln} N_e(\zeta_i,{\bm a})\over\partial a} \to 
   \int {\rm d}\zeta\, {N_t(\zeta)\over N_e(\zeta,{\bm a})} 
   {\partial N_e(\zeta,{\bm a})\over\partial a} .  
\end{displaymath}
Maximizing the likelihood means that we vary ${\bm a}$ until both 
these expressions are equal.  

Suppose that the true distribution would produce $N_t(\lambda,\zeta)$, 
and that this true distribution is well described by a particular 
choice of the parameters, say ${\bm a}_t$.  Then the question is, 
are the two expressions equal when ${\bm a}={\bm a}_t$?  If not, our 
definition of $p_i$ is incorrect, because the minimum will occur at 
some other value of ${\bm a}$.  To see that it is the correct choice, 
note that when 
$N_e(\lambda,\zeta,{\bm a}_t) = N_t(\lambda,\zeta)$, then the first 
expression becomes 
\begin{displaymath}
  \int {\rm d}\zeta\,\int {\rm d}\lambda 
             {\partial N_e(\lambda,\zeta,{\bm a}_t)\over\partial a} = 
  \int {\rm d}\zeta\,{\partial N_e(\zeta,{\bm a}_t)\over\partial a}.  
\end{displaymath}
And because 
\begin{displaymath}
 N_e(\zeta,{\bm a}_t) = \int {\rm d}\lambda\,N_e(\lambda,\zeta,{\bm a}_t) 
               = \int {\rm d}\lambda\,N_t(\lambda,\zeta) = N_t(\zeta),
\end{displaymath}
the second expression also reduces to 
 $\int {\rm d}\zeta\,{\partial N_e(\zeta,{\bm a})/\partial a}$.  
Thus, both the sums over $i$ reduce to the same quantity.  
Hence, maximizing the expression for the likelihood given above 
(equation~\ref{mlphotoz}) does indeed yield an accurate unbiased 
estimate of the luminosity function.  
This also demonstrates that omission of the ${\rm d}V_{\rm com}$ terms 
present in our expression for $p_i$ would lead to a biased estimate 
of the shape of the luminosity function.

\section{Some applications}\label{apps}

\subsection{Galaxies and QSOs:  $dN/dz$ and $\phi(L)$}  
The methods above allow one to reconstruct the intrinsic $dN/dz$ and 
$\phi(L)$ distributions of, e.g., QSOs, LRGs and other galaxy 
distributions in, e.g., the SDSS.  These will be useful for a number 
of clustering analyses, as well as for studying galaxy evolution.  
As a proof of concept, Figure~\ref{nbc} shows the result of running 
the $dN/dz$ deconvolution algorithm on publically available data.  
The input QSO catalog is from application of the Non-parametric 
Bayesian Classification algorithm to the SDSS DR1: 
this produced a catalog of about 100,000 objects (Richards et al. 2004).
For each object, photometric redshifts were determined following 
Weinstein et al. (2004).  
About 22,000 of these objects have spectra from which a spectroscopic 
redshift estimate is available.  For this subset of objects, the top 
left panel compares $z_{\rm phot}$ and $z_{\rm spec}$.  
The other panels show that the distribution of 
$p(z_{\rm phot}-z_{\rm spec}|z_{\rm spec})$ is rather complex.  
The panels on the right show the differences between the 
true- (filled circles) and photo-$z$ distributions (histograms), 
and that the deconvolution algorithm (curve) does a reasonable job 
reconstructing the former from the latter.  

\subsection{Peculiar velocities}

\begin{figure}
 \centering
 \includegraphics[width=\hsize]{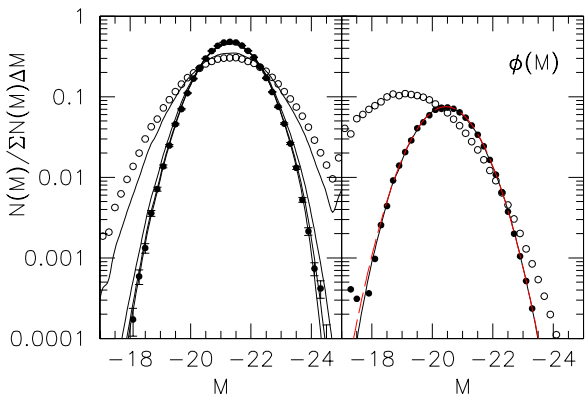}
 \vspace{-3cm}
 \caption{Same as Figure~\ref{NMvpec}, but now the underlying luminosity 
          function is Lognormal (Gaussian in absolute magnitude), and 
          the errors in distance are also assumed to be Lognormal.   
          Parameters were chosen to mimic early-type galaxies in the 
          SDSS (from Bernardi et al. 2003), and distance errors were 
          chosen to be about a factor of two larger than in typical 
          peculiar velocity surveys.  
          Notice how the raw photo-$z$ estimate of $N(M)$ (open 
          circles in panel on left) is broader than the true 
          distribution (filled circles), making the estimated 
          luminosity function have a slight excess of luminous 
          galaxies, and a significantly larger excess of faint 
          galaxies (open circles in panel on right).  
          Nevertheless, when started from the estimated distribution, 
          our deconvolution algorithm quickly converges to the true 
          distribution.}
 \label{NMgaussian}
\end{figure}

In peculiar velocity surveys such as SFI, ENEAR, EFAR and 6dF, the 
distance indicator (the Tully-Fisher, $D_n-\sigma$, or Fundamental 
Plane relations) is noisy:  typically this noise is approximately 
twenty percent of the distance, or about 0.4~mags.  If uncorrected for, 
a generic effect of distance uncertainties is to inflate the estimated 
number of low (and high) luminosity galaxies.  This is illustrated in 
Figure~\ref{NMgaussian}, where the error has been set to 1~mag so 
that the effect is more clearly seen.  
Since the faint end of the luminosity function provides a strong 
constraint on galaxy formation models, it is important that it be 
measured accurately.  Therefore, it may be interesting to apply 
our methods to data from peculiar velocity surveys.  In particular, 
such methods may be necessary for estimating unbiased luminosity 
functions from HIPASS and ALFALFA.  

\subsection{The stellar luminosity function}
Distances to stars are sometimes estimated by the method of photometric 
parallax:  essentially, this method uses the offset from a color 
magnitude-relation to infer a distance.  Because the color-magnitude 
relation almost certainly has intrinsic scatter (current estimates 
are about 0.5~mags), the associated distance estimate is noisy:  this 
is entirely analogous to the noise in distance estimates from peculiar 
velocity surveys.  
Determination of the stellar luminosity function is an important 
ingredient in understanding the IMF.  
Most current determinations are based on the method of 
Stobie et al. (1989) which assumes small errors in the distance 
estimate, and requires prior knowledge of the shape of the luminosity 
function.  
Since our non-parametric methods are accurate even when the noise on 
the distance estimate is large, it may be interesting to apply our 
methods to this problem as well.  

\subsection{Correlations between observables}
So far, I have mainly discussed how to make accurate estimates of the 
true redshift and luminosity distributions when only noisy distance 
estimates are available.  However, noisy distance estimates have 
another important effect for which it is possible to correct.  
Namely, galaxy observables are known to correlate with one-another:  
the most well-known of these is the correlation between luminosity $L$ 
and circular velocity $V_c$ (the Tully-Fisher relation for spirals), 
or velocity dispersion $\sigma$ (the Faber-Jackson relation for ellipticals), 
but physical size $R$ and surface brightness $I$ are also well-correlated 
(the Kormendy relation for ellipticals), and $L$ also correlates with 
size and color for both spirals and ellipticals.  
Every one of these correlations has been used to constrain galaxy 
formation models, and, since one measures apparent magnitudes and 
angular sizes rather than absolute magnitudes and physical sizes, every 
one of these relations includes at least one distance-dependent quantity.  

\begin{figure}
 \centering
 \includegraphics[width=\hsize]{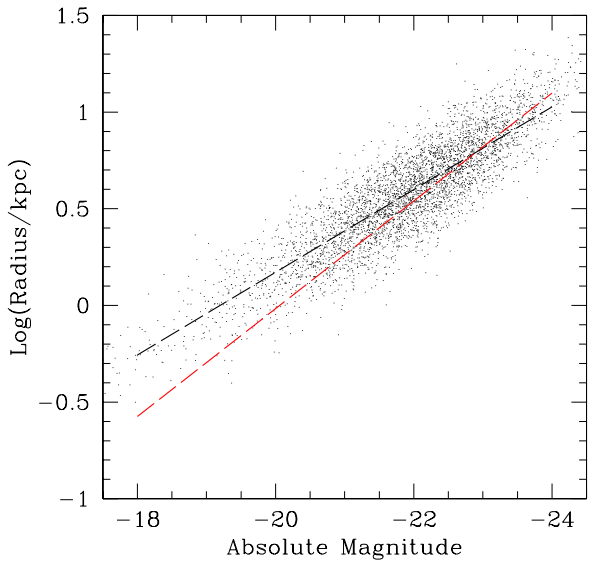}
 \caption{Effect of distance errors on the estimated correlation 
          between size and luminosity for early-type galaxies.  
          Steeper dashed line shows the true $\langle\sigma|L\rangle$ 
          relation, and shallower line shows a least squares fit to the 
          dots which were obtained by using the photo-$z$ estimate of the 
          distance to compute the sizes and absolute magnitudes.  } 
 \label{LRphotoz}
\end{figure}

Noise in the distance estimate will lead to biased estimates of these 
correlations.  To illustrate, Figure~\ref{LRphotoz} shows the correlation 
between luminosity and size in a catalog which is constructed to mimic 
the SDSS early-type galaxy sample (Bernardi et al. 2003).  
The steeper dashed curve shows the true $\langle R|L\rangle$ relation.  
The dots show the result of assuming that 
 $z_{\rm phot} = z_{\rm spec}\ + $\ gaussian with rms 0.03 
(this amount of scatter in the distance estimate is realistic), 
and then recomputing the absolute magnitude and size using 
$z_{\rm phot}$ instead of $z_{\rm spec}$.  The shallower dashed line 
shows $\langle R_{\rm phot}|L_{\rm phot}\rangle$:  the change in slope 
is dramatic.

The qualitative nature of the effect is easy to understand.  
Distance errors scatter objects towards the bright and faint luminosity 
tails.  This increases the spread along the absolute magnitude axis.  
If this were the only effect, then one might expect the $R-L$ relation 
to be shallower.  However, the distance error causes a correlated change 
to the size:  assuming an object is closer than it really is makes one 
infer a smaller luminosity {\em and size} than it really has.  So the 
net motion of each point is left-and-down, or right-and-up.  
If these motions were parallel to the principal axis of the true relation, 
the net effect would only be to change the scatter of the relation.  
In this case, they are not, so small distance errors have a non-negligible 
effect.  

This bias is simpler to correct-for when only one of the 
variables is distance-dependent.  For instance, in the case of the 
$L$-color relation, it is only $L$ which is affected by the distance 
error (this is not quite true, because $k$-corrections depend on 
wavelength---I am mainly using this to illustrate an argument).  
This suggests that if the distance indicator is unbiased in the mean, 
then the mean $L$ as a function color can be estimated directly.  
(The scatter around this mean relation is interesting in its own right:  
it will, of course, be affected by the noise in the distance estimate.)  
In practice, however, even this case is not entirely straightforward, 
because galaxy catalogs are almost always magnitude limited, and this 
introduces selection effects into the estimate of 
$\langle L|{\rm color}\rangle$; absent distance errors, it is 
$\langle{\rm color}|L\rangle$ rather than $\langle L|{\rm color}\rangle$ 
which can be estimated free of selection effects!  
When accurate distances are known, these selection effects can be 
accounted for by using the quantity $V_{\rm max}$ which played an 
important role in our discussion of the luminosity function.  This 
suggests that the methods discussed previously should allow one to 
estimate such correlations in photometric galaxy catalogs.  
When the distance error appears in both variables, it is slightly 
harder to correct, but a correction is still possible.  Essentially, 
one simply needs to write the expressions given previously in 
matrix rather than scalar notation.  Making this generalization 
correctly is the subject of work in progress.

\section{Discussion}

I presented two algorithms for estimating the intrinsic redshift 
and luminosity distributions from photo-$z$ surveys.  
These algorithms improve on previous work by Subbarao et al. (1996) 
and Chen et al. (2003).  Subbarao et al. concluded that numerical 
simulations were necessary to derive accurate estimates---my 
analysis shows that simulations can be avoided.  
Chen et al. wrote down a maximum likelihood expression 
which they then maximized---I find a different expression for 
the likelihood, and provide an analytic calculation which shows 
that maximizing this expression does indeed lead to an unbiased 
estimate; maximizing their expression instead would return a biased 
answer.  

The error in the photometric redshift gives rise to an error in the 
estimated luminosity.  Since measurement errors in the apparent 
magnitude also give rise to errors in the estimated luminosity, it is 
tempting to treat the photo-$z$ errors similarly to how one treats 
the effects of errors in the photometry.  However, the two errors are 
not equivalent for the simple but important reason that the photo-$z$ 
error, while affecting the estimated luminosity, leaves the observed 
apparent magnitude unchanged.  In this respect, it is more accurate to 
view the photo-$z$ error as equivalent to a peculiar velocity.  This 
motivates reanalysis of relatively shallow galaxy surveys for which the 
peculiar velocity may be a substantial fraction of the observed redshift, 
e.g. faint, nearby, low surface brightness galaxies, 
     or galaxies in the 6dF survey (Jones et al. 2004).  
In this case, the error in the true distance comes from the thickness 
of the Fundamental Plane, or the $D_n-\sigma$ relation, and is 
typically on the order of twenty percent.  

The fractional error on the distances to most stars in our galaxy 
(those for which parallax measurements are not available) is relatively 
large.  Stobie, Ishada \& Peacock (1989) discuss a method for 
estimating the luminosity function in the case of photometric parallaxes 
derived from the color-magnitude relation, but the approach is 
parametric (it requires an accurate guess of the intrinsic shape of 
the luminosity function), and it assumes that the distance errors are 
small.  Our approach provides accurate non-parametric estimates which 
are valid even when the errors are large.  We intend to apply our methods 
to provide non-parametric estimates of the stellar luminosity function 
which are not compromised by the noise in the distance estimator.  

Both the maximum likelihood and the $V_{\rm max}$ estimators I derived 
assume that galaxies do not evolve, so one must break the sample up 
into narrow redshift bins before analysis.  
This is risky in principle, because one wants a narrow bin in true 
redshift, but only photo-$z$s are available.  In practice, photo-$z$s 
are sufficiently accurate that a narrow bin in photo-$z$ is still 
quite narrow in true-$z$.
The maximum-likelihood and $V_{\rm max}$ estimators of the luminosity 
function have another drawback:  they ignore the fact that different 
galaxy types require different $k(z)$-corrections, so one must preselect 
the sample to insure that it contains galaxies that are of the same 
type.  Extending the analysis to allow for evolution and type is clearly 
desirable, and is the subject of work in progress.  

\section*{Acknowledgments}
This work was completed while I was at the University of Pittsburgh 
where L. Lucy had developed his deconvolution algorithm some thirty 
years previously.  
I thank Adrian Collister and Gordon Richards for supplying ANN$z$ 
and NBC data in 2003, Bridget Falck and Sam Schmidt for their 
patience while the $V_{\rm max}$ and maximum likelihood algorithms 
were being developed, Donald Lynden-Bell for a discussion about the 
history of the $V_{\rm max}$ method, and the Aspen Center for Physics 
where this work, which is supported by NSF Grant 0520677, was finally 
written-up.

\label{lastpage}
\end{document}